\documentclass[aps,prd,preprint,longbibliography]{revtex4-1}

\usepackage{graphicx}
\usepackage{amsmath}
\usepackage{amsfonts}
\usepackage{natbib}
\usepackage{comment}

\newcommand\fft[2]{\frac{#1}{#2}}
\newcommand\nn{\nonumber}

\begin{document}

\preprint{MCTP-16-31}

\title{Spatial Anisotropy in Nonrelativistic Holography}

\author{Joshua W. Foster}
\affiliation{Indiana University, Bloomington, IN 47405}
\affiliation{Michigan Center for Theoretical Physics, Randall Laboratory of Physics, The University of Michigan, Ann Arbor, MI 48109}

\author{James T. Liu}
\affiliation{Michigan Center for Theoretical Physics, Randall Laboratory of Physics, The University of Michigan, Ann Arbor, MI 48109}

\date{\today}

\begin{abstract}
We examine holographic theories where Lifshitz symmetry is broken with spatial anisotropy.  In particular, we focus on the conditions imposed by the null energy condition, and demonstrate that it is possible to have unusual anisotropic fixed points where a subset of the spatial dimensions have negative scaling exponents.  We also construct interpolating solutions between UV and IR fixed points and show that there is essentially no restriction placed on the endpoints of the flow once anisotropic scaling is allowed.  As an example, we demonstrate a flow from AdS to AdS with $c_{IR}>c_{UV}$ that is allowed by the null energy condition since it proceeds through an intermediate Lorentz-violating region. Finally, we examine the holographic Green's function in anisotropic Lifshitz spacetimes.
\end{abstract}

\maketitle

\section{Introduction}

In the relativistic context, AdS/CFT has by now been remarkably well established as a powerful tool
for addressing the strongly coupled regime of quantum field theories.  While the most direct
application is to superconformal field theories, holography can provide insights on a large class of
systems.  A prime example is the investigation of the strongly coupled quark-gluon plasma
and the holographic computation of the ratio of the shear viscosity to entropy density.

Following the successes of relativistic holography, there has been much recent interest in
constructing gravitational backgrounds dual to non-relativistic theories and investigating their
properties.  Perhaps the most straightforward example is that of Lifshitz scaling
\begin{equation}
t\to\lambda^z t,\qquad \vec x\to\lambda\vec x,
\end{equation}
where $z$ is the dynamical exponent.  This symmetry can be holographically realized by introducing
a radial coordinate $r$ and corresponding bulk metric of the form \cite{Kachru:2008yh}
\begin{equation}
ds^2=-\fft{dt^2}{r^{2z}}+\fft{d\vec x^2+dr^2}{r^2}.
\end{equation}
It is also possible to extend the Lifshitz symmetry to the full Galilean conformal group, in which
case the holographic dual has two extra dimensions 
\cite{Son:2008ye,Balasubramanian:2008dm}.

While Lifshitz scaling governs the behavior of a wide range of condensed matter systems
near quantum critical points, there has also been recent interest in the holographic description
of systems that break rotational invariance.  For example, one may consider a system in the
nematic phase, where spatial rotation is broken by long-range directional order.  In such cases,
the critical point would be described by spatially anisotropic scaling
\begin{equation}
t\to\lambda^z t,\qquad x_i\to\lambda^{p_i}x_i,
\label{eq:sas}
\end{equation}
with a corresponding gravitational dual
\begin{equation}
ds^2=-\fft{dt^2}{r^{2z}}+\left(\sum_i\fft{dx_i^2}{r^{2p_i}}\right)+\fft{dr^2}{r^2}.
\label{eq:alif}
\end{equation}
This background can be realized in a model with multiple massive vector fields
\cite{Taylor:2008tg}.  Lifshitz domain wall solutions with spatial anisotropy have been considered in
\cite{Azeyanagi:2009pr,Cremonini:2014pca,Roychowdhury:2015cva,Roychowdhury:2015fxf}.

In the case of spatially isotropic Lifshitz symmetry, imposition of the null energy
condition in the holographic dual gives rise to the condition $z\ge1$ for the dynamical
exponent.  The $z=1$ case corresponds to relativistic scaling, while $z=2$ arises,
for example, in the quantum Lifshitz model \cite{Ardonne2004} and the quadratic band crossing model
\cite{Sun2009}.  We may expect a similar condition to hold for spatially anisotropic systems.  Note,
however, that by rescaling the parameter $\lambda\to\lambda^{1/z}$ in (\ref{eq:sas}), it
is easy to see that only the ratios $p_i/z$ are physical.  This suggests the generalization of
$z\ge1$ to be $z\ge\max\{p_i\}$, which we confirm below using the null energy condition
in the bulk.

More generally, we wish to investigate the allowed values of $\{z,p_i\}$ so we can classify
the set of models with anisotropic Lifshitz scaling.  We use as our guiding principle the null
energy condition in the bulk, and find both lower and upper limits on the scaling exponents $\{p_i\}$.
The upper limit is simply $p_i\le z$ for all $i$, which is equivalent to the lower bound on $z$
given above.  The lower limit, however, is rather curious, in that it is possible for some but not
all of the $\{p_i\}$ to be negative without violating the null energy condition.  This is a
rather unusual situation for holography, as negative exponents correspond to directions
opening up in the IR, as opposed to the UV.  If all exponents, including $z$, were negative,
then we would simply interchange what we call the UV and the IR.  However, the case here
corresponds to simultaneous positive and negative exponents, so having directions that can
open up in the IR is unavoidable.

From a holographic point of view, a background with a negative scaling exponent is somewhat
unusual. While the overall volume of the radial slices is increasing towards the UV, the negative exponent directions are instead shrinking.  In a sense, this is similar to the Kasner metric, where at least one direction must shrink while the others expand.  This make the holographic interpretation more problematic as the usual connection between UV asymptotics and scaling dimensions of the dual operators breaks down for modes associated with the shrinking dimensions.  Furthermore, the UV geometry has tidal singularities that are reminiscent of Lifshitz singularities at the horizon.

While these difficulties arise in a pure anisotropic Lifshitz bulk with negative scaling, physically what would be more realistic is a model with conventional UV asymptotics, such as AdS or spatially isotropic Lifshitz, flowing into a spatially anisotropic IR.  In this case, any negative exponents will potentially lead to unusual IR physics, but would not lead to any drastic modifications to the
boundary asymptotics.  Nevertheless, this is a rather unusual situation for holography.  If we trace a direction with negative scaling exponent in the IR, then it opens up both in the UV and the IR.  In the spatially isotropic case, this would correspond to a domain wall flow from UV to UV, which is forbidden by the holographic $c$-theorem, or, equivalently, the null energy condition.  Removing the requirement of spatial isotropy, the best we can do is have a subset of directions opening up in both sides of the flow.  Although we do not construct such backgrounds
explicitly, we will show that it is possible to interpolate between positive and negative exponents
in both directions while satisfying the null energy condition.  Flows from `horizon to horizon' would also be interesting to consider, as they may share some features with the Randall-Sundrum scenario.

In section~\ref{sec:nec}, we explore the consequences of the null energy condition in the bulk.  We first identify restrictions on the scaling exponents $\{z,p_i\}$ at fixed points and then turn to interpolation geometries that flow between scaling regions in the UV and the IR.  We comment on the possible realization of negative exponent geometries in vector-scalar models coupled to gravity and also demonstrate that the tidal singularity at the Lifshitz horizon persists in the anisotropic case. In Section~\ref{sec:green} we explore the holographic Green's function in models with anisotropic scaling.  In particular, we make use of the WKB approximation to demonstrate universal features of the spectral function in the presence of spatial anisotropy.  Finally, we conclude in Section~\ref{sec:conc} with a toy example of an interpolating geometry connecting AdS to AdS with $L_{UV}<L_{IR}$.

\section{Spatially Anisotropic Backgrounds}
\label{sec:nec}

As indicated above, we are interested in translationally invariant systems with anisotropic scale invariance.  The corresponding bulk geometry can be described by a metric of the form
\begin{equation}
    ds^2_{n+2}=-e^{2A}dt^2+\left(\sum_{i=1}^ne^{2B_i}dx_i^2\right)+dr^2,
    \label{eq:ABmet}
\end{equation}
where we have introduced the functions $A(r)$ and $B_i(r)$.  At a scale invariant fixed point the functions will take the form
\begin{equation}
    A=\fft{zr}L,\qquad B_i=\fft{p_ir}L,
    \label{eq:fpcond}
\end{equation}
where the critical exponents are given by the constants $\{z,p_i\}$, and $L$ sets the curvature scale in the bulk.  This is a fairly general class of geometries, and it can be seen that with suitable selection of parameters, we may achieve well-studied geometries such as AdS spacetimes ($z=p_i=1$) or Lifshitz spacetimes with critical exponent $z$ ($p_i=1$). The spatially anisotropic geometries of interest in this work are those with at least some subset of $p_i \neq p_j$.

\subsection{Conditions on the scaling exponents}

Far from being arbitrary parameters of our geometry, values of $\{z, p_i\}$ are constrained by the requirement that our geometries be solutions of Einstein's equations supported by physically reasonable mass-energy distributions. The null energy condition provides a relatively simple means to impose such a constraint. Through Einstein's equation, we will write the null energy condition in terms of the Ricci tensor $R_{\mu \nu}$ and arbitrary null vector $\xi_\mu$  as $R_{\mu \nu}\xi^\mu \xi^\nu \geq 0$. 

For the metric (\ref{eq:ABmet}), a straightforward computation of the Ricci tensor gives
\begin{align}
    R_{tt}&=-g_{tt}\Bigl[A''+A'\Bigl(A'+\sum_iB_i'\Bigr)\Bigr]\to-\fft{g_{tt}}{L^2}\Bigl[z\Bigl(z+\sum_ip_i\Bigr)\Bigr],\nn\\
    R_{ij}&=-g_{ij}\Bigl[B_i''+B_i'\Bigl(A'+\sum_kB_k'\Bigr)\Bigr]\to-\fft{g_{ij}}{L^2}\Bigl[p_i\Bigl(z+\sum_kp_k\Bigr)\Bigr],\nn\\
    R_{rr}&=-\Bigl[A''+A'^2+\sum_i\Bigl(B_i''+B_i'^2\Bigr)\Bigr]\to-\fft1{L^2}\Bigl[z^2+\sum_ip_i^2\Bigr],
    \label{eq:Riccis}
\end{align}
where the expressions on the right correspond to the scale invariant fixed point given by (\ref{eq:fpcond}). Applying the null energy condition in the $t$-$x_i$ direction gives
\begin{equation}
    (z-p_i)\Bigl(z+\sum_jp_j\Bigr)\ge0,
    \label{eq:txi}
\end{equation}
and in the $t$-$r$ direction gives

\begin{equation}
    \sum_{i=1}^n\Bigl(z p_i- p_i^2\Bigr) \geq 0,
    \label{eq:tr}
\end{equation}
where $n$ is the number of spatial field theory directions. Here and subsequently, we choose $z>0$ since we may always make the substitution $r\to\frac{1}{r}$ such that this is true. Applying the Cauchy-Schwarz inequality to (\ref{eq:tr}), we find $n z \geq \sum_i p_i$. Then summing over index $i$ in (\ref{eq:txi}), we have $(nz - \sum_i p_i)(z + \sum_j p_j) \geq 0$. The term $(nz - \sum_i p_i)$ may be positive or vanishing, but in either case, we may conclude that $(z + \sum_j p_j)$ is positive and that therefore, from (\ref{eq:txi}), that $z-p_i \geq 0$ for all $i$.

\begin{figure}[t]
    \centering
    \includegraphics[width=6cm]{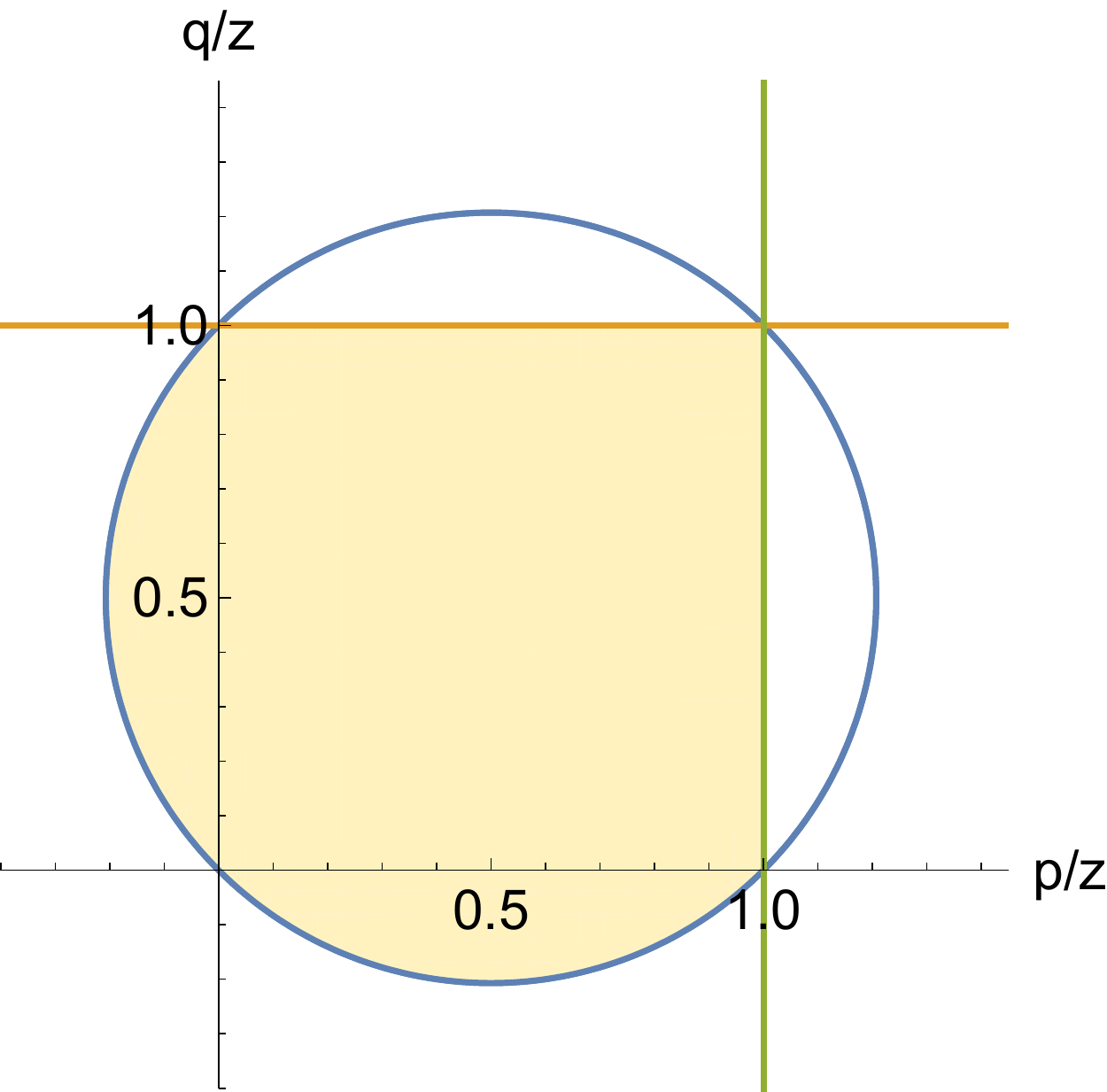}
    \caption{For the $2+1$ dimensional case, the allowed values of $\{z,p,q\}$ are shown as the shaded region in the $p$-$q$ plane.}
    \label{fig:nec}
\end{figure}

The condition in (\ref{eq:tr}) may also be nicely rearranged to
\begin{equation}
    \sum_{i=1}^n\Bigl(p_i-\fft{z}2\Bigr)^2\le \fft{nz^2}4,
    \label{eq:cir}
\end{equation}
showing that for a given value of $z$, the $\{p_i\}$ lie inside a sphere of radius $z\sqrt{n}/2$ centered at $p_i=z/2$. As an example, the allowed values of $\{z,p,q\}$ in the $2+1$ dimensional case is shown in Fig.~\ref{fig:nec}.

Several interesting features of the constraints from the null energy condition may be observed in this figure. First, for given rotationally invariant systems, we have $p=q$, and the allowed parameters then lie along the diagonal in Fig.~\ref{fig:nec}, with the origin corresponding to $z=\infty$ and the upper right corresponding to the relativistic case $z=1$.  Second, in the anisotropic case, while $p$ and $q$ cannot simultaneously be negative, each may individually attain negative values without violation of the null energy condition.  We will further investigate the possibility of having a negative scaling exponent as it gives rise to unusual behavior in the AdS/CFT correspondence.

\subsection{Interpolating geometries}

What we have seen above is that at a scale invariant fixed point, the conditions imposed on the scaling exponents by the null energy condition are
\begin{equation}
p_i\le z\quad(i=1,2,\ldots,n),\qquad
\sum_{i=1}^n\Bigl(p_i-\fft{z}2\Bigr)^2\le \fft{nz^2}4.
\end{equation}
However, more generally, we are interested in bulk geometries that may flow from one fixed point to another.  In particular, it would be natural to consider a spacetime which displays Lorentz symmetry in the UV but flows to a Lifshitz geometry in the IR.

Even without considering a complete bulk model, we can in principle consider a background that interpolates between UV and IR metrics by choosing a function $f(r)$ which varies smoothly from $0$ in the IR to $1$ in the UV.
We can then empirically write down a bulk metric $g_{\mu\nu}$ that interpolates between the IR and UV.  One natural choice is \cite{Kachru:2013voa}
\begin{equation}
g_{\mu\nu}= \left[1-f(r)\right]g^{IR}_{\mu\nu} + f(r)g^{UV}_{\mu\nu}.
\end{equation}
Just as in our previous analysis, we may seek to ensure that this metric obeys the null energy condition. If we restrict $g^{IR}$ and $g^{UV}$ to be metrics which individually obey the null energy condition, the existence of a satisfactory $f(r)$ such that $g_{\mu\nu}$ obeys the null energy condition seems likely. However, this is challenging to show in general.

Alternatively, for the Lifshitz geometries described by (\ref{eq:ABmet}), we may choose to interpolate the metric functions $A(r)$ and $B_i(r)$
\begin{align}
    A(r)&=f(r)A^{UV}(r)+(1-f(r))A^{IR}(r),\nn\\
    B_i(r)&=f(r)B_i^{UV}(r)+(1-f(r))B_i^{IR}(r).
    \label{eq:interp}
\end{align}
The reason this remains manageable is that the Ricci components (\ref{eq:Riccis}) are straightforward combinations of derivatives of $A(r)$ and $B_i(r)$.  Assuming $A^{UV/IR}$ and $B_i^{UV/IR}$ take the form (\ref{eq:fpcond}), we then have
\begin{align}
A'&=g\frac{z_{UV}}{L_{UV}}+(1-g)\frac{z_{IR}}{L_{IR}},&
A''=g'\left(\frac{z_{UV}}{L_{UV}}-\frac{z_{IR}}{L_{IR}}\right),\nn\\
B_i'&=g\frac{p_{i\,UV}}{L_{UV}}+(1-g)\frac{p_{i\,IR}}{L_{IR}},&
B_i''=g'\left(\frac{p_{i\,UV}}{L_{UV}}-\frac{p_{i\,IR}}{L_{IR}}\right),
\end{align}
where $g\equiv (rf)'$.

Although $f(r)$ directly interpolates the functions $A$ and $B$, we see that $g(r)$ and its derivative are what appears in the Ricci tensor.  We demand both $f$ and
$g$ to be monotonic from $0$ in the IR to $1$ in the UV.  For example, we may choose
\begin{equation}
f=\fft12\left(1+\fft{r_0}r\log\cosh\fft{r}{r_0}\right),\qquad g=\fft12\left(1+\tanh\fft{r}{r_0}\right),\qquad
g'=\fft1{2r_0}\mbox{sech}^2\fft{r}{r_0},
\end{equation}
where $r_0$ sets the width of the interpolation region.

The null energy condition in the $t$-$x_i$ direction is equivalent to $-R_t^t+R_i^i\ge0$, where
\begin{align}
-R_t^t+R_i^i&=(A-B_i)''+(A-B_i)'(A+\sum_jB_j)'\nonumber\\
&=g'(\alpha_{i\,UV}-\alpha_{i\,IR})+[g\alpha_{i\,UV}+(1-g)\alpha_{i\,IR}][g\beta_{UV}+(1-g)\beta_{IR}].
\label{eq:NECi}
\end{align}
Here we have defined
\begin{equation}
\alpha_i=\fft{z-p_i}{L},\qquad\beta=\fft{z+\sum_{i=1}^np_i}{L}.
\end{equation}
Since we assume the UV and IR geometries both satisfy the null energy condtion, we have $\alpha_i\ge0$ and $\beta>0$.  Since $g(r)$ is monotonic between $0$ and $1$, the second term in (\ref{eq:NECi}) is non-negative, and we may write
\begin{equation}
-R_t^t+R_i^i\ge g'(\alpha_{i\,UV}-\alpha_{i\,IR})+\min(\alpha_{i\,UV},\alpha_{i\,IR})\min(\beta_{UV},\beta_{IR}).
\end{equation}
If $\alpha_{i\,UV}$ and $\alpha_{i\,IR}$ are both non-vanishing, then $\min(\alpha_{i\,UV},\alpha_{i\,IR})$ is strictly positive, and we can always satisfy the $t$-$x_i$ null energy condition by making $g'$ arbitrarily small [e.g.\ by letting $r_0\to\infty$ in (\ref{eq:interp})]. If either $\alpha_{i\,UV}=0$ or $\alpha_{i\,IR}=0$, then we can return to (\ref{eq:NECi}) to show that the null energy condition can be satisfied for sufficiently slowly varying $g$.  Finally, if $\alpha_{i\,UV}=\alpha_{i\,IR}=0$, then $-R_t^t+R_i^i=0$, and the null energy condition is
trivially satisfied.  This demonstrates that the $t$-$x_i$ null energy condition can always be satisfied with appropriate choice of $f(r)$ for the solution (\ref{eq:interp}) interpolating between arbitrary scaling geometries in the UV and IR.

Of course, we also have to consider the null energy condition in the $t$-$r$ direction. Here we find
\begin{align}
-R_t^t+R_r^r&=\fft{g^2}{L_{UV}^2}\sum_{i=1}^n\left[\left(\fft{z_{UV}}2\right)^2\!\!-\left(p_{i\,UV}-\fft{z_{UV}}2\right)^2\right]\!
+\fft{(1-g)^2}{L_{IR}^2}\sum_{i=1}^n\left[\left(\fft{z_{IR}}2\right)^2\!\!-\left(p_{i\,IR}-\fft{z_{IR}}2\right)^2\right]\nonumber\\
&\quad+2\fft{g(1-g)}{L_{UV}L_{IR}}\sum_{i=1}^n\left[\left(\fft{z_{UV}}2\right)\left(\fft{z_{IR}}2\right)
-\left(p_{i\,UV}-\fft{z_{UV}}2\right)\left(p_{i\,IR}-\fft{z_{IR}}2\right)\right]\nonumber\\
&\quad-g'\sum_{i=1}^n\left(\fft{p_{i\,UV}}{L_{UV}}-\fft{p_{i\,IR}}{L_{IR}}\right).
\label{eq:NECr}
\end{align}
Assuming the UV and IR exponents both satisfy the `circle' condition (\ref{eq:tr}), then the first line is automatically
non-negative.  The second line is then non-negative by the Cauchy-Schwarz inequality.  The third line can have
either sign, depending on the sum of exponents $p_i/L$ in the UV and IR.  However, it can be made
arbitrarily small by choosing a sufficiently slowly varying $g$, so that we can ensure that $-R_t^t+R_r^r\ge0$ along the entire flow, and hence that the null energy condition is satisfied.  There is, however, one exception to this argument, and that is when the first two lines in (\ref{eq:NECr}) are identically zero.  This only happens if both UV and IR scaling exponents lie on the circle and are parallel, $\{z_{UV},p_{i\,UV}\}\parallel\{z_{IR},p_{i\,IR}\}$.  Without loss of generality, we choose $\{z_{UV},p_{i\,UV}\}=\{z_{IR},p_{i\,IR}\}$, in which case (\ref{eq:NECr}) reduces to
\begin{equation}
-R_t^t+R_r^r=\left(g'\sum_{i=1}^np_i\right)\left(\fft1{L_{IR}}-\fft1{L_{UV}}\right).
\end{equation}
The null energy condition is then satisfied for
$L_{UV}\ge L_{IR}$.  This essentially reproduces the relativistic $c$-theorem for the case $z=p_i=1$, but also holds for the negative exponent solutions lying, e.g., on the circle in Fig.~\ref{fig:nec}.

Except for the special case with fixed $\{z,p_i\}$ on the circle, we see that it is always possible to choose an interpolating function $f(r)$ so that the null energy condition is satisfied everywhere along the flow between arbitrary UV and IR scaling solutions, so long as the UV and IR fixed points themselves satisfy the null energy condition.
While we have only considered flows between two fixed points, we can interpolate between multiple scaling regions by patching together flows.  One way this may arise is, e.g., in a flow from AdS$_{n+2}$ in the UV to Lifshitz in an intermediate region and finally to AdS$_2\times\mathbb R^n$ in the deep IR.

Finally, it is worth emphasizing that since the UV and IR data are (almost) arbitrary, flows are reversible in that the UV and IR solutions can be swapped without violating the null energy condition.  Moreover, flows from AdS to AdS can have $L_{UV}<L_{IR}$, in apparent violation of the relativistic $c$-theorem, provided they proceed through an intermediate Lorentz-violating region.

\subsection{Constructing Holographically Dual Models}
While one might anticipate that the restriction to spacetimes compatible with the null energy condition might be equivalent to the restriction to physically realizable spacetimes, it is unclear if this is always true. Therefore, while the null energy condition provides a convenient means of pruning the space of solutions to the field equations, the explicit construction of field configurations that lead to the realization of a given geometry is a valuable exercise. In this section, we will neglect the construction of flowing geometries as even the construction of scaling geometries will be seen to be sufficiently challenging. 

The construction of holographically dual models with spatial anisotropy was investigated by Taylor in \cite{Taylor:2008tg}. In Taylor's construction, geometries of the form (\ref{eq:alif}) are realized subject to the condition that $z \geq p_1 \geq p_2 \hdots \geq p_n \geq 0$. Although this is a fairly comprehensive result, it does not include the negative parameter geometries seemingly allowed by the null energy condition, raising the question of if they can be realized by a suitable field configuration.

We begin by considering a system with a scalar and vector field without any interaction with Lagrangian density
\begin{equation}
    \mathcal{L}_M = -\frac{1}{4}F_{\mu \nu}F^{\mu \nu}-\frac{1}{2}M^2 A^\mu A_\mu-\frac{1}{2}\partial_\mu \phi \partial^\mu \phi - \frac{1}{2} m^2 \phi^2.
\label{eq:onefield}
\end{equation}
We work with the anisotropic Lifshitz metric given in the form (\ref{eq:alif}) and consider the form of the bulk Einstein equation.  In particular, we examine the combination $T^{0}_{0} - T^{\mu}_{\nu} = R^{0}_{0} - R^{\mu}_{\nu}$ relevant to the null energy condition.
We are specifically interested in the cases $\mu = \nu = i \neq r$ and $\mu = \nu = r$. For convenience, we will take $M \rightarrow 0$. As will soon be shown, this loss of generality is acceptable. For the first case we find 
\begin{equation}
    g^{00}\sum_{j \neq i}g^{jj}(F_{0j})^2 - g^{ii}\sum_{j \neq i}g^{jj}(F_{i j})^2 + g^{00}(\partial_0 \phi)^2 - g^{ii}(\partial_i \phi)^2 = (p_i-z)(z+\sum_j p_j).
\end{equation}
In the second case, we have
\begin{equation}
g^{00}\sum_{j \neq r}g^{jj}(F_{0j})^2 - g^{rr}\sum_{j \neq r}g^{jj}(F_{r j})^2 + g^{00}(\partial_0 \phi)^2 - g^{rr}(\partial_i \phi)^2 = \sum_j(p_j^2 - z p_j).
\end{equation}
The left side of each equation may be recognized as negative semi-definite, so we conclude that $(z-p_i)(z+\sum_j p_j)  \geq 0$ and $\sum_j(z p_j -p_j^2) \geq 0$. 

This is of course a restatement of the null energy condition in the particular model with a massive scalar field and massless vector field.  This is a very simple case of the more general scenario with arbitrary numbers of massive scalar and vector fields, and it therefore seems at least plausible that a suitable field configuration might realize a negative parameter geometry as allowed by the null energy condition.

Nonetheless, it remains highly challenging, if even possible, to construct an explicit example of such a field configuration. It might initially seem promising to build a realizable geometry with massive vector fields through the construction of \cite{Taylor:2008tg} and then to perturb it to a negative parameter geometry through the addition of scalar fields. By construction, the massive vector fields will result in a constant-valued stress energy tensor $T^\mu_\nu$, and we aim to maintain constant-valuedness after the perturbation by the scalar fields. We would begin with a Lagrangian generalizing (\ref{eq:onefield}) to multiple fields. However, computing stress-energy tensor $T^\mu_\nu$ and comparing the sums of differences of every possible pairing of its diagonal elements, which in turn must be constant, we see that arbitrary sums and differences of $g^{\mu \mu}\sum_i(\partial_\mu \phi^{(i)})^2$ and $g^{\nu \nu}\sum_i(\partial_\nu \phi^{(i)})^2$ are constant. Then each of $g^{\mu \mu}\sum_i(\partial_\mu \phi^{(i)})^2$ themselves must be constant. Then by the positive-semidefiniteness of each term in the sum, $\partial_\mu \phi^{(i)} = 0$ or $\partial_\mu \phi^{(i)} \propto \sqrt{\| g_{\mu \mu}\|}$. Due to the anisotropy, this result is contradictory unless $\phi^{(i)}$ is only allowed to possibly depend on $r$. Then $\partial_r\phi^{(i)} = {A^{(i)}}/{r}$ so $\phi = A^{(i)}\log(r)$. But even for a freely chosen mass, such a $\phi^{(i)}$ cannot satisfy an independent Klein-Gordon equation unless $A^{(i)} = 0$ so that the field is vanishing and no perturbations are made to the vector-field-generated geometry.

With this highly limiting result in mind, it is clear that if a negative parameter geometry is possible to construct, its construction will require dispensing of the assumption that the vector fields produce a constant-valued contribution to the stress-energy tensor, thereby allowing for non-constant contributions to the stress-energy tensor by the scalar fields. We might also try allowing the vector fields to be massive, but this is essentially equivalent to allowing for nonconstant-valued contributions to the stress-energy tensor by the vector fields since we will have to consider terms like $F_{\mu\nu}^{(i)2}$ and $A_{\mu}^{(i)2}$, which enter the Lagrangian and therefore the stress-energy tensor with differing powers of metric factors. The overall effect of either approach is to require the full consideration of the combined Einstein and matter equations.
Disentangling these PDEs is generically difficult, and it remains to be seen if a negative parameter geometry can be realized in a simple model.

\subsection{Singularities}

Before turning to the holographic Green's function, we recall that Lifshitz spacetimes are characterized by a tidal singularity at the horizon \cite{Kachru:2008yh,Hartnoll:2009sz,Copsey:2010ya,Horowitz:2011gh}. As this is related to the breaking of relativistic invariance, we expect that the horizon singularity would have a similar structure in the spatially anisotropic case.  To see this, we follow \cite{Horowitz:2011gh}, and consider a radial timelike geodesic specified by $t(\tau)$ and $r(\tau)$ in the metric (\ref{eq:alif}).  Defining the conserved energy $E=\dot t/r^{2z}$, where the dot indicates differentiation with respect to proper time $\tau$, the radial equation is then
\begin{equation}
    \dot r^2 = E^2r^2\left(1-\fft1{E^2r^{2z}}\right).
\end{equation}
This suggests that we define the parallelly propagated frame
\begin{align}
     (e_0)^\mu &= Er^{2z}\fft\partial{\partial t}+Er^{z+1}\sqrt{1-\fft1{E^2r^{2z}}}\fft\partial{\partial r},\nn\\
     (e_r)^\mu &= Er^{2z}\sqrt{1-\fft1{E^2r^{2z}}}\fft\partial{\partial t}+Er^{z+1}\fft\partial{\partial r},\nn\\
     (e_i)^\mu &= r^{p_i}\fft\partial{\partial x^i}.
\end{align}
A straightforward calculation then gives the non-vanishing Riemann components
\begin{align}
R_{0r0r} &= z^2,\nn\\
R_{0i0i} &= p_i^2+p_i(z-p_i)E^2r^{2z},\nn\\
R_{0iri} &= p_i(z-p_i)E^2r^{2z}\sqrt{1-\fft1{E^2r^{2z}}},\nn\\
R_{riri} &= -zp_i+p_i(z-p_i)E^2r^{2z},\nn\\
R_{ijij} &= -p_ip_j\qquad(i\ne j).
\end{align}
(No sums are to be taken over $i$ and $j$.)

In this coordinate system, the horizon is located at $r\to\infty$.  Hence we see that the components $R_{0i0i}$, $R_{0iri}$ and $R_{riri}$ all diverge at the horizon whenever $z\ne p_i$.  As a result, the tidal singularity is generic to spatially anisotropic Lifshitz spacetimes.  The only exception is the AdS case when $z=p_i$ for all $i$.  Note, however, that if some subset of $p_i$'s are equal to $z$, then the tidal forces diverge only along the directions orthogonal to this subset.  Furthermore, the divergence is present even in directions with negative exponent $p_i$ that expand towards the horizon.

\section{The holographic Greens function}
\label{sec:green}

In order to explore some of the consequences of spatial anisotropy, we may investigate the holographic Green's function in such backgrounds.  For a relativistic conformal field theory, the scalar Green's function in momentum space takes the simple form
\begin{equation}
    G(q^2)\sim (-q^2)^{\Delta-d/2},
\end{equation}
where $\Delta$ is the conformal dimension of the corresponding operator and $d$ is the dimension of the CFT.  For isotropic Lifshitz scaling, on the other hand, the Green's function has additional freedom, and may be written as
\begin{equation}
    G(\omega,\vec k)\sim |\vec k|^{2\nu z}\mathcal G(\omega/|\vec k|^z).
\end{equation}
The Green's function was obtained analytically in \cite{Kachru:2008yh} for $z=2$ Lifshitz, and can be studied numerically or via WKB approximation for other values of $z$.  At high frequencies, the Green's function behaves as a power law, $G\sim\omega^{2\nu}$, while at low frequencies there is exponential suppression of spectral weight \cite{Faulkner:2010tq,Hartnoll:2012wm,Keeler:2014lia,Hartnoll:2012rj}.

Turning to the anisotropic Lifsitz case, we expect a similar power-law scaling at high frequencies coupled to exponential suppression at low frequencies.  However, additional energy scales are involved, depending on the components $k_i$ of the transverse momenta.  To see this behavior more explicitly, consider a bulk scalar with action
\begin{equation} 
S = \int d^4x \sqrt{-g}\left[-\frac{1}{2}g^{\mu \nu}\partial_{\mu}\phi \partial_{\nu} \phi - \frac{1}{2}m^2 \phi^2\right].
\end{equation}
Working with the spatially anisotropic metric (\ref{eq:alif}) and making use of translational symmetry, we look for a solution of the form
\begin{equation}
\phi(\omega,\vec k, r) = f(r)e^{i(\vec k\cdot\vec x-\omega t)}.
\end{equation}
The bulk scalar equation of motion then becomes
\begin{equation}
-r^2 f'' + (z-1+\sum p_i)rf' +\left(m^2-\omega^2 r^{2z} + \sum k_i^2 r^{2p_i}\right)f = 0. 
\end{equation}
We find it convenient to make the change of variables
\begin{equation}
r=\rho^{1/z},\qquad f(r)=\rho^{\fft12\sum\bar p_i}\psi(\rho).
\end{equation}
This brings the scalar equation into the Schr\"odinger form
\begin{equation}
-\psi''(\rho)+V(\rho)\psi(\rho)=\bar\omega^2\psi(\rho),
\label{eq:schr}
\end{equation}
where
\begin{equation}
V(\rho)=\fft{\nu^2-1/4}{\rho^2}+\sum\fft{\bar k_i^2}{\rho^{2(1-\bar p_i)}}.
\label{eq:schpot}
\end{equation}
Here we have defined the rescaled quantities
\begin{equation}
\bar\omega=\fft\omega{z},\qquad\bar k_i=\fft{k_i}z,\qquad\bar p_i=\fft{p_i}z
\end{equation}
and the energy scaling exponent
\begin{equation}
\nu=\sqrt{\left(\fft{m}z\right)^2+\left(\fft{1+\sum\bar p_i}2\right)^2}.
\end{equation}

\begin{figure}[t]
\centering
    \includegraphics[width=10cm]{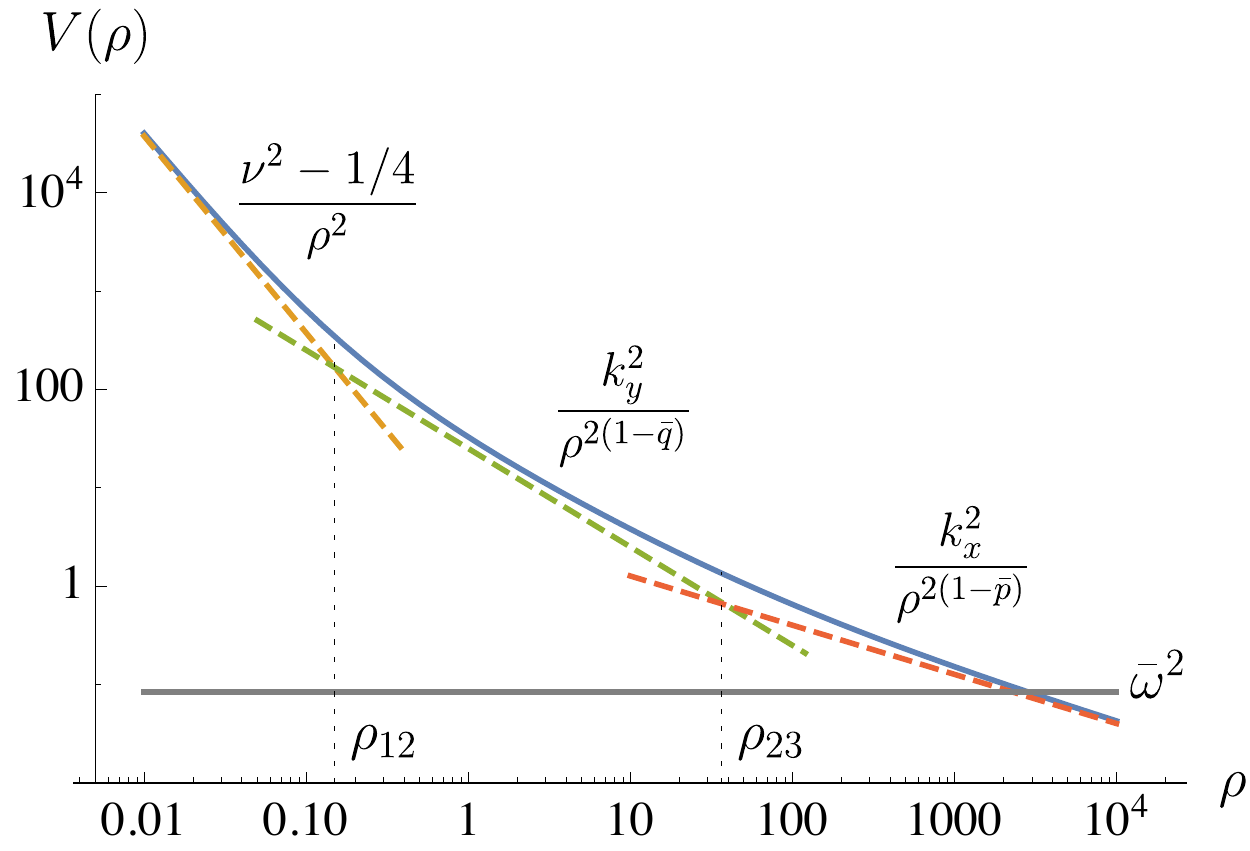}
    \caption{Effective potential with three scaling regions.  An example value of $\hat\omega$ is shown where the classical turning point is in the final scaling region.}
    \label{fig:Veff}
\end{figure}

One advantage of the Schr\"odinger form (\ref{eq:schr}) is that it is intuitively straightforward to understand the behavior of the solutions based on the potential (\ref{eq:schpot}).  Since we impose $\bar p_i\le1$ by the null energy condition, $V(\rho)$ is a sum of negative powers of $\rho$.  For simplicity, we now focus on the $2+1$ dimensional case, and take the scaling exponents $\{z,p,q\}$ ordered according to $z>p>q$.  (This system captures all the essential features of spatial anisotropy, and its generalization to higher dimensions is straightforward.)  The explicit potential is then
\begin{equation}
V(\rho)=\fft{\nu^2-1/4}{\rho^2}+\fft{\bar k_y^2}{\rho^{2(1-\bar q)}}+\fft{\bar k_x^2}{\rho^{2(1-\bar p)}}.
\label{eq:Vpq}
\end{equation}
So long as $\bar q>0$, the three terms in the potential are ordered by the magnitude of their exponents; the first term dominates in the UV ($\rho\to0$), while the last term dominates in the IR ($\rho\to\infty$).  An example of this potential is given in Figure~\ref{fig:Veff}.

Given the Schr\"odinger potential (\ref{eq:schpot}), the solution to the radial equation near the boundary takes the form
\begin{equation}
f(\rho)=A(\rho^{\Delta_-}+\cdots)+B(\rho^{\Delta_+}+\cdots),
\end{equation}
where
\begin{equation}
\Delta_\pm=\fft{1+\sum\bar p_i}2\pm\nu.
\end{equation}
The coefficients $A$ and $B$ of the non-normalizable and normalizable modes, respectively, are generally functions of $\omega$ and $\vec k$.  Following the prescription of \cite{Son:2002sd}, the holographic Green's function is then given by
\begin{equation}
G(\omega,\vec k)=K\fft{B(\omega,\vec k)}{A(\omega,\vec k)},
\end{equation}
where $K$ is a normalization constant.  The retarded Green's function is obtained by taking infalling boundary conditions at the horizon.

There is one subtlety to this prescription, however, and that is that we have assumed the dominant boundary behavior of the Schr\"odinger potential is $V(\rho)\sim1/\rho^2$.  This holds for $z>p>q>0$, but recall that negative exponents are allowed by the null energy condition.  Whenever $\bar q<0$, the second term in (\ref{eq:Vpq}) in fact dominates (assuming $\bar k_y\ne0$), and the potential is more singular than $1/\rho^2$ at the boundary.
In this case, the radial behavior is no longer power-law, but instead takes the form
\begin{equation}
f(\rho)\sim e^{\pm\bar k_y\rho^{\bar q}/\bar q}\rho^{(1-\bar q)/2}.
\end{equation}
In this case, the usual prescription for defining the holographic Green functions fails. Physically, this is due to the exchanged roles of $\rho \rightarrow 0$ as the horizon and $\rho \rightarrow \infty$ as the boundary whenever the corresponding Lifshitz exponent is negative. Under such an interchange, imposing boundary conditions at $\rho \rightarrow \infty$ and extracting the Green's function at $\rho\to0$ is no longer appropriate.

In order to make sense of geometries with a negative Lifshitz exponent is to restrict them to the IR region in the bulk and to connect them to a more conventional UV geometry, whether AdS or Lifshitz.  In this case, the corresponding negative exponent direction will have little effect on the IR physics, as its contribution to the Schr\"odinger potential will fall off rapidly near the horizon.  Although it would be interesting to explore such geometries, we will not pursue it further here.  Thus, from now on, we will restrict ourselves to the case with all positive exponents, $0<\bar q<\bar p<1$.

\subsection{WKB Approximation}

In general, the radial equation (\ref{eq:schr}) does not admit elementary solutions, as the potential (\ref{eq:Vpq}) has non-integral exponents in $\rho$.  Of course, explicit solutions are available when $\bar p$ and $\bar q$ are either $0$, $1/2$ or $1$.  (This covers the relativistic and $z=2$ Lifshitz cases.)  However, even in the absence of an analytical solution, we can determine the general features of the holographic Green's function by examining the effective potential in Figure~\ref{fig:Veff}.

The potential $V(\rho)$ is the sum of three power-law terms, and can be broken up into regions where each term is dominant.  We let $\rho_{12}$ denote the transition between the first and second, and $\rho_{23}$ the transition between the second and third scaling regions.  We then have
\begin{equation}
\rho_{12}\approx\left(\fft\nu{\bar k_y}\right)^{1/\bar q},\qquad
\rho_{23}\approx\left(\fft{\bar k_y}{\bar k_x}\right)^{1/(\bar p-\bar q)}.
\end{equation}
In order to have three separate regions, we demand that $\rho_{23}>\rho_{12}$, which is equivalent to
\begin{equation}
\left(\fft{\bar k_y}\nu\right)^{\bar p}>\left(\fft{\bar k_x}\nu\right)^{\bar q}\qquad\mbox{for}\quad V(\rho)\approx\fft{\nu^2-1/4}{\rho^2}\to\fft{\bar k_y^2}{\rho^{2(1-\bar q)}}\to\fft{\bar k_x^2}{\rho^{2(1-\bar p)}}.
\label{eq:three}
\end{equation}
In the opposite case, we transition directly from the first to the third scaling region
\begin{equation}
\left(\fft{\bar k_y}\nu\right)^{\bar p}<\left(\fft{\bar k_x}\nu\right)^{\bar q}\qquad\mbox{for}\quad V(\rho)\approx\fft{\nu^2-1/4}{\rho^2}\to\fft{\bar k_x^2}{\rho^{2(1-\bar p)}}.
\label{eq:two}
\end{equation}

In either the two or three region case, the general features of the holographic Green's function can be read off from the effective Schr\"odinger potential (\ref{eq:Vpq}).  For fixed spatial momentum, at high frequencies, the classical turning point is in the $1/\rho^2$ region, and the resulting Green's function is of power-law form
\begin{equation}
    G_R(\omega,\vec k)\sim \omega^{2\nu}\quad\mbox{as}\quad\omega\to\infty.
    \label{eq:GRpower}
\end{equation}
At low frequencies, however, the classical turning point is in a shallower region of the potential, and the radial wavefunction corresponds to a quantum mechanical tunneling problem.  In this case, the holographic Green's function will approach a real constant (at fixed $\vec k$).  In particular, it exhibits an exponential suppression of spectral weight \cite{Faulkner:2010tq,Hartnoll:2012wm,Keeler:2014lia}
\begin{equation}
\chi(\omega,\vec k)=2\,\mbox{Im}\,G_R(\omega,\vec k)\sim\exp\left[-\fft{\sqrt\pi\Gamma\left(\fft{\bar p}{2(1-\bar p)}\right)}{z\Gamma\left(\fft1{2(1-\bar p)}\right)}\left(\fft{k_x}{\omega^{\bar p}}\right)^{\fft1{1-\bar p}}\right]
\qquad\mbox{as}\quad\omega\to0,
\label{eq:sfact}
\end{equation}
so long as the classical turning point is in the right-most region of the potential shown in Figure~\ref{fig:Veff}.  In the two-region case, the spectral function then interpolates between power-law (\ref{eq:GRpower}) and suppression (\ref{eq:sfact}).  In the three-region case, the spectral weight gets suppressed in the intermediate region as well, ending up with a maximum suppression of (\ref{eq:sfact}) in the limit of sufficiently small $\omega$.

The behavior alluded to above can be quantified using a WKB approximation \cite{Faulkner:2010tq, Keeler:2013msa,Keeler:2014lia}.  While this approximation is unreliable for the real part of the holographic Green's function, the imaginary part remains under control.  The resulting expression for the spectral function is
\begin{equation}
\chi(\omega,\vec k)=\lim_{\epsilon\to0}\epsilon^{-2\nu}e^{-2S_{\vec k}(\epsilon,\omega)},
\end{equation}
where
\begin{equation}
S_{\vec k}(\epsilon,\omega)=\int_\epsilon^{\rho_0}d\rho\sqrt{\hat V(\rho)-\bar\omega^2}.
\label{eq:integrand}
\end{equation}
Here $\rho_0$ is the classical turning point, $\hat V(\rho_0)=\bar\omega^2$, and $\hat V(\rho)$ is given by (\ref{eq:Vpq}) with the shift $\nu^2\to\nu^2+1/4$ (see e.g.~\cite{Keeler:2013msa}).  However, even in this case, the WKB integral does not have a closed form solution for generic parameters.  Nevertheless, we can make some further approximations for certain limiting cases.

We first consider the two-region case given by (\ref{eq:two}).  Physically, this corresponds to taking $\bar k_y$ sufficiently small, so that the retarded Green's function is only probing the $x$-direction.  The spectral function then takes on a power-law form at high frequencies and becomes exponentially suppressed at low frequencies.  The behavior matches the spatially isotropic case, so we have
\begin{align}
\chi(\omega,k_x,k_y)\approx\begin{cases}
\left(\fft{e\bar k_x}{2\nu}\right)^{2\nu/\bar p}\exp\left[-\fft{\sqrt\pi\Gamma\left(\fft{\bar p}{2(1-\bar p)}\right)}{\Gamma\left(\fft1{2(1-\bar p)}\right)}\left(\fft{\bar k_x}{\bar\omega^{\bar p}}\right)^{\fft1{1-\bar p}}\right],&\fft{\bar\omega}\nu<\left(\fft{\bar k_x}\nu\right)^{1/\bar p};\\
\left(\strut\fft{e\bar\omega}{2\nu}\right)^{2\nu},&\fft{\bar\omega}\nu>\left(\fft{\bar k_x}\nu\right)^{1/\bar p}.
\end{cases}
\end{align}

We now shift our attention to the three-region case given in (\ref{eq:three}).  Here, both $k_x$ and $k_y$ are important, and as the frequency is lowered, the spectral weight is first suppressed by $y$-momentum, and then by $x$-momentum, as can be seen in Figure~\ref{fig:Veff}.  (This ordering is because we have chosen $p>q$).
When the classical turning point is to the left of $\rho_{23}$, the $x$-direction is irrelevant, and the spectral function takes the standard asymptotic form.  When the classical turning point is in the final scaling region, we can break up the WKB integral into three terms and perform a matched asymptotic expansion.  The result is
\begin{align}
\chi(\omega,k_x,k_y)\approx\begin{cases}
\left(\fft{e\bar k_y}{2\nu}\right)^{2\nu/\bar q}\exp\left[
\fft{\Gamma\left(\fft{-\bar p}{2(\bar p-\bar q)}\right)\Gamma\left(\fft{\bar q}{2(\bar p-\bar q)}\right)}{2\sqrt\pi(\bar p-\bar q)}\left(\fft{\bar k_y^{\bar p}}{\bar k_x^{\bar q}}\right)^{\fft1{\bar p-\bar q}}\right.\\
\kern7em\left.-\fft{\sqrt\pi\Gamma\left(\fft{\bar p}{2(1-\bar p)}\right)}{\Gamma\left(\fft1{2(1-\bar p)}\right)}\left(\fft{\bar k_x}{\bar\omega^{\bar p}}\right)^{\fft1{1-\bar p}}\right],&\fft{\bar\omega}\nu<\fft1\nu\left(\fft{\bar k_x^{1-\bar q}}{\bar k_y^{1-\bar p}}\right)^{\fft1{\bar p-\bar q}};\\
\left(\fft{e\bar k_y}{2\nu}\right)^{2\nu/\bar q}\exp\left[-\fft{\sqrt\pi\Gamma\left(\fft{\bar q}{2(1-\bar q)}\right)}{\Gamma\left(\fft1{2(1-\bar q)}\right)}\left(\fft{\bar k_y}{\bar\omega^{\bar q}}\right)^{\fft1{1-\bar q}}\right],&\fft1\nu\left(\fft{\bar k_x^{1-\bar q}}{\bar k_y^{1-\bar p}}\right)^{\fft1{\bar p-\bar q}}<\fft{\bar\omega}\nu<\left(\fft{\bar k_y}\nu\right)^{1/\bar q};\\
\left(\strut\fft{e\bar\omega}{2\nu}\right)^{2\nu},&\fft{\bar\omega}\nu>\left(\fft{\bar k_y}\nu\right)^{1/\bar q}.
\end{cases}
\end{align}
Strictly speaking, these expressions are only valid when there is a large separation of scales between $k_x$ and $k_y$ and when the frequency is not close to any of the transition points.  Nevertheless, the general behavior, from power-law to exponential suppression, is apparent.

\section{Conclusion}
\label{sec:conc}

In this paper, we have conducted some initial efforts toward the systematic study of general spatially anisotropic Lifshitz backgrounds in holography. The main focus of our analysis as been a classification of allowed backgrounds consistent with the application of the null energy condition.  We find a large class of allowable geometries including, somewhat surprisingly, a class of ``negative exponent geometries," where some subset of dimensions close in the UV and open up in the IR due to a negative scaling exponent. We attempted to explicitly construct field configurations which support these negative exponent geometries as solutions to Einstein's equations, but found the task generically difficult and incompatible with existing construction techniques.

We further examined geometries that interpolate between scale-invariant fixed points in the UV and the IR.  Here we find that the null energy condition in itself places very few restrictions on the allowed fixed points that can be connected by a bulk interpolating geometry.  In particular, it is always possible to interpolate between arbitrary UV and IR scaling geometries, except for the special case of scaling exponents lying on the circle of Fig.~\ref{fig:nec}.

As mentioned above, it is in fact possible to construct a flow from AdS in the UV to AdS in the IR with $L_{UV}<L_{IR}$, in apparent violation of the relativistic $c$-theorem \cite{Alvarez:1998wr,Girardello:1998pd,Freedman:1999gp,Sahakian:1999bd,Myers:2010xs,Myers:2010tj}, so long as the flow proceeds through an intermediate Lifshitz region.  An example of such a flow is given by the bulk metric
\begin{equation}
    ds^2=-\left(e^{r/L_{UV}}+\fft{L_{UV}}{L_{IR}}e^{r/L_{IR}}\right)^2dt^2+\left(e^{r/L_{UV}}+e^{r/L_{IR}}\right)^2d\vec x^2+dr^2,
\label{eq:flowMetric}
\end{equation}
This flow is engineered so that it interpolates between the UV and IR metrics
\begin{align}
    ds_{UV}^2&=e^{2r/L_{UV}}(-dt^2+d\vec x^2)+dr^2,&&r\to\infty,\nn\\
    ds_{IR}^2&=e^{2r/L_{IR}}(-(L_{UV}/L_{IR})^2dt^2+d\vec x^2)+dr^2,&&r\to-\infty.
\end{align}
(Note that we demand $L_{UV}<L_{IR}$, so that $e^{r/L_{UV}}$ dominates in the UV.)  While both regions are asymptotically AdS, we see that the IR geometry picks up a redshift factor $z=L_{IR}/L_{UV}-1$ compared with the UV.

\begin{figure}[t]
  \includegraphics[width=10cm]{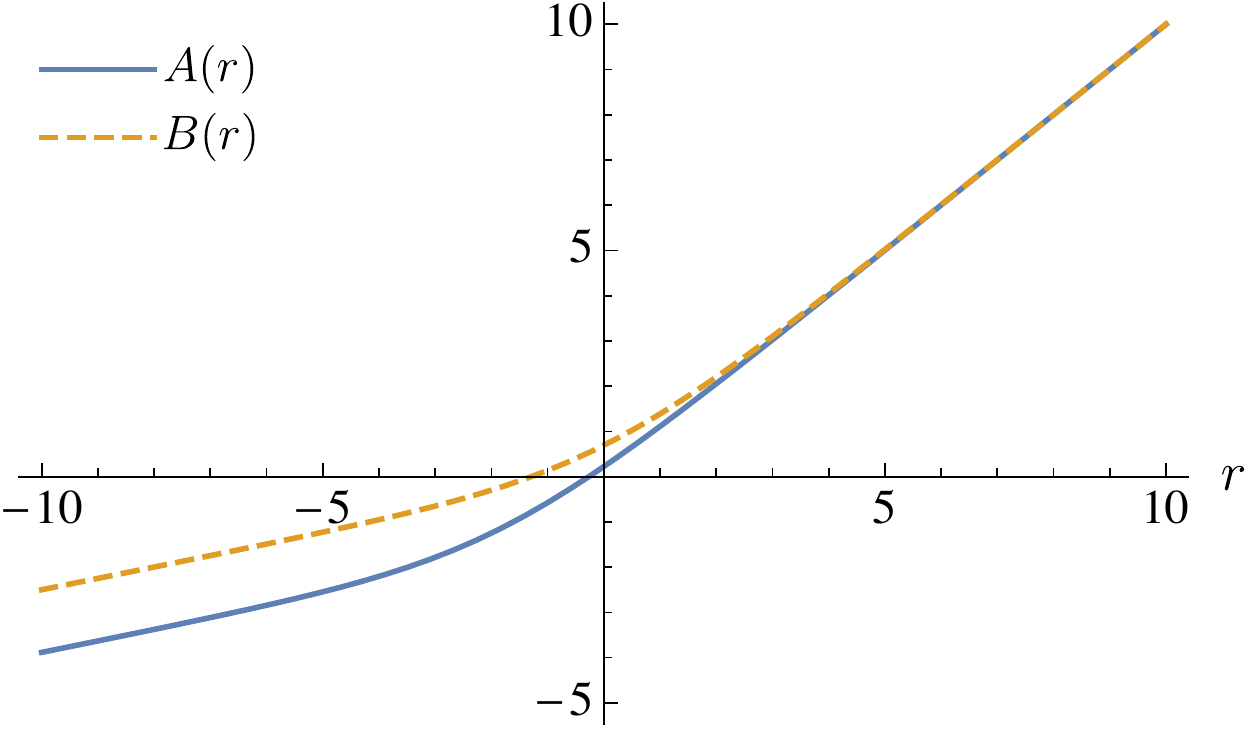}
  \caption{The functions $A(r)$ and $B(r)$ for the AdS to AdS interpolating metric (\ref{eq:flowMetric}), where we have chosen $L_{UV}=1$ and $L_{IR}=4$.}
\label{fig:ABflow}
\end{figure}

This flow is of course not a true violation of the holographic $c$-theorem, as the assumption of Lorentz invariance is broken along the flow.  As shown in Fig.~\ref{fig:ABflow}, the metric functions diverge along the flow, with $B(r)$ first attaining its IR form, followed by $A(r)$.  The intermediate region is a spatially anisotropic
Lifshitz geometry with critical exponent $z_{\textrm{Lif}}=L_{IR}/L_{UV}$.  An explicit IIB supergravity
flow from AdS$_5$ to AdS$_5$ with $L_{UV}=L_{IR}$ has recently been constructed in \cite{Donos:2016zpf}.
This flow proceeds through an intermediate Lifshitz-like region with scaling exponents  $\{1,1,1,2/3\}$
\cite{Azeyanagi:2009pr}.  (See also \cite{Chesler:2013qla,Donos:2014gya} for examples of AdS to AdS
flows with $L_{IR}=L_{UV}$ that proceed through intermediate regions with broken Lorentz invariance.)
It would be interesting to see whether any realistic bulk theory can lead to flows with $L_{IR}$ strictly
greater than $L_{UV}$, and hence with $c_{IR}>c_{UV}$.

\section*{Acknowledgements}

JF thanks Pranav Rao for stimulating discussion. This material is based upon work supported by the National Science Foundation under Grant No.~PHY~1262543 and by the US Department of Energy under Grant No.~DE-SC0007859.

\bibliography{bibliography}

\end{document}